\documentclass[page-classic]{epl2} 

\usepackage{amsmath}
\usepackage{color}
\usepackage{ulem}

\newcommand{\marge}[1]{\marginpar{}}  % do not show margin notes
\newcommand{\Sl}[1]{{}}           % do not show labels
\newcommand{\beq}[1]{\Sl{#1}\begin{equation}\if#1\empty\else\label{#1}\fi}
\newcommand{\eeq}{\end{equation}}
\newcommand{\beqa}[1]{\Sl{#1}\begin{eqnarray}\if#1\empty\else\label{#1}\fi}
\newcommand{\eeqa}{\end{eqnarray}}

\newcommand{\Int}{\displaystyle \int}

\newcommand{\Ex}[1]{(\ref{#1})}

\shorttitle{Nonextensive thermodynamics and classical Hamiltonian systems}
\institute{
 \inst{1} Center for Nonlinear Phenomena and Complex Systems CP 231 \\
  Universit\'{e} Libre de Bruxelles, 1050 Brussels, Belgium 
}

\begin{document}

\title{Questioning the validity of non-extensive \\
thermodynamics for classical Hamiltonian systems }
\author{James F. Lutsko$^1$\thanks{E-mail: \email{jlutsko@ulb.ac.be}} and 
{Jean Pierre Boon$^1$\thanks{E-mail: \email{jpboon@ulb.ac.be}}}}
\shortauthor{ J.F. Lutsko  and J.P. Boon}


%{\color{red} Text}

\abstract{We examine the non-extensive approach to the statistical mechanics of
Hamiltonian systems with $H=T+V$ where $T$ is the classical kinetic energy.
Our analysis starts from the basics of the formalism by applying the
standard variational method for maximizing the entropy subject to the
average energy and normalization constraints. The analytical results show
(i)~that the non-extensive thermodynamics formalism should be called into
question to explain experimental results described by extended exponential
distributions exhibiting long tails, i.e. $q$-exponentials with $q>1$, and
(ii)~that in the thermodynamic limit the theory is only consistent in the
range $0\leq q\leq1$ where the distribution has finite support, thus
implying that configurations with e.g. energy above some limit have zero
probability, which is at variance with the physics of systems in contact
with a heat reservoir. We also discuss the ($q$-dependent) thermodynamic
temperature and the generalized specific heat.}
 
\maketitle

\section{Introduction}

The observation of natural phenomena and of
laboratory experiments provides a wide spectrum of experimental results
showing distributions of data that deviate from exponential decay as
predicted for Boltzmann behavior~\cite{davis}. It was the goal of
non-extensive statistical mechanics developed originally by Tsallis~\cite%
{tsallis88} to offer a new approach to explain the non-Boltzmann behavior of
non-equilibrium systems \footnote{We note that other approaches exist such 
as super-statistics \cite{beck-cohen} and the fractional and nonlinear 
Fokker-Planck equations  (see e.g. \cite{lutsko-boon} and references therein)}. 
More precisely the primary motivation for
non-extensive thermodynamics is as a way to understand deformed exponential
distributions (such as $q$-exponentials exhibiting long tails when $q>1$)
found empirically in many areas of physics and other scientific disciplines~%
\cite{swinney-tsallis}. The interest raised by this new approach has grown
over the years and has produced an abundant literature~\cite{tsallis09}
reflecting new theoretical developments and a considerable number of
applications to subjects as diverse as defect turbulence, energy
distribution in cosmic rays, earthquake magnitude value distributions and
velocity distributions in micro-organism populations or distributions of
financial market data. Parallel to these developments, critical analyses
were presented questioning the merits of the non-extensive theory~\cite%
{opponents} and thereby of its applications as well. These criticisms raised
questions that often gave rise to ontological conflicts~\cite{conflict}. 
%between proponents and skeptics of the theory~\cite{conflict}.
Most of the criticisms are based on phenomenological analyses and
thermodynamic arguments questioning the compatibility of the theory with
classical statistical thermodynamics. More recently, the non-extensive
theory was also re-examined on the basis of analyses demonstrating the
necessary {{discreteness}} of systems to which the theory applies~\cite{abe_10}
and the limits of validity of the non-extensive formalism for a Hamiltonian
system, the $q$-ideal gas, a model system of independent quasi-particles
where the interactions are incorporated in the particles properties~{{\cite%
{abe_99, boon_lutsko_11}}}. Here we merely adopt the viewpoint of analytical rigor to
establish the validity limits of the non-extensive formalism for the general
class of classical Hamiltonian systems with continuous variables and
consequently of the class of physical systems to which the non-extensive
interpretation applies.

In the formalism, the $q$-exponential distributions arise as the result of a
variational calculation maximizing the generalized entropy, the $q$-entropy,
under the constraints of normalizability of the distribution function and of
a prescribed average energy~\cite{tsallis09}. The goal of the present work
is to investigate the exact form of the distribution so derived for
classical Hamiltonian systems from both the usual Tsallis entropy~\cite%
{tsallis88} and the homogeneous entropy~\cite{lutsko_boon_grosfils}. The
conclusion is found to be the same in both cases: the theory is only
consistent in the thermodynamic limit for $0\leq q\leq1$. For finite systems
of $N$ particles, the upper limit is $1+a/N$ for $a\sim{\mathcal{O}}(1)$.
This means (i)~that the non-extensive thermodynamics formalism cannot be
used, at least in any straightforward way, to explain phenomena for which one
observes that $q$ takes a value $q>1$ {{(corresponding to asymptotically power-law 
distributions)}} i.e. when the distributions exhibit extended exponential
forms with long tails, and (ii)~that, within the validity domain, the
distribution has finite support, thus implying that configurations with e.g.
energy above some limit have zero probability, a strange situation for
systems with typical molecular potentials which are steeply repulsive at
small distances. We treat successively the case according to the development
based on the Tsallis entropy~$S_{q}$ and the case based on the homogeneous
entropy~$S^{H}_{q}$. {{The reason for examining the two cases is the criterion of 
stability against perturbations of the probability distribution function, or
{\it {Lesche stability}}~\cite{lesche_abe}}}  : the
homogeneous entropy was shown to be Lesche stable while the Tsallis entropy 
is not~\cite{lutsko_boon_grosfils}. We also present the results for the generalized 
($q$-dependent) thermodynamic properties of Hamiltonian systems in both cases.

\section{Tsallis entropy}

 Non-extensive statistical mechanics is developed
on the basis of three axioms: (i)~the $q$-entropy for systems with
continuous variables is given by~\cite{tsallis09} 
\begin{equation}
S_{q}\,=\,k_{B}\,\frac{1-{K\int \rho ^{q}\left( \Gamma \right) d\Gamma }}{q-1%
}\,,  \label{STq}
\end{equation}%
where $\Gamma$ is the phase space variable and
$K$ must be a quantity with the dimensions of $\left[ \Gamma \right]
^{q-1}$, i.e. $K=\hbar ^{DN\left( q-1\right) }$
($D$ denotes the dimension of the system and $N$ the number of particles)
and the classical Boltzmann-Gibbs entropy is retrieved in the limit $q\rightarrow 1$; 
(ii)~the distribution function $\rho \left( \Gamma \right) $ is slaved to the
normalization condition 
\begin{equation}
1\,=\,\int \rho \left( \Gamma \right) d\Gamma \,;  \label{Norm}
\end{equation}%
(iii)~the internal energy is measured as 
\begin{equation}
U\,=\,{\int P_{q}\left( \Gamma \right) H\,d\Gamma }\,=\,\frac{\int \rho
^{q}\left( \Gamma \right) H\,d\Gamma }{\int \rho ^{q}\left( \Gamma \right)
d\Gamma }\,,  \label{Uesc}
\end{equation}%
where $P_{q}\left( \Gamma \right) $ is the escort probability distribution~%
\cite{beck_schlogl} which is the actual probability measure. We consider
Hamiltonian systems with $H=T+V$ where $T$ is the classical kinetic energy.
The distribution function $\rho \left( \Gamma \right) $ is obtained by
maximizing the $q$-entropy subject to the normalization (\ref{Norm}) and
average energy (\ref{Uesc}).
%constraints using Lagrange multipliers to obtain 
Introducing the Lagrange multipliers $\bar{\alpha}$ and $\bar{\gamma}$, the variational method  leads to
\begin{equation}
0 = \frac{k_{B}Kq}{1-q}\rho^{q-1}(\Gamma)-\bar{\alpha} - q\bar{\gamma}\frac{(H-U)}{\int \rho ^{q}\left( \Gamma \right)d\Gamma }\rho^{q-1}(\Gamma) \,,
\label{vareq}
\end{equation}
which is solved to yield 
\begin{equation}
\rho \left( \Gamma \right) =\frac{\left( 1-\left( 1-q\right) \gamma \left(
H-U\right) \right) ^{\frac{1}{1-q}}}{\displaystyle\int \left( 1-\left(
1-q\right) \gamma \left( H-U\right) \right) ^{\frac{1}{1-q}}d\Gamma }\,,
\label{rhoT1}
\end{equation}%
with  $\gamma = \frac{\bar{\gamma}}{K\int \rho ^{q}\left( \Gamma \right)d\Gamma }$ and
where the normalization condition (\ref{Norm}) has been used to eliminate the
multiplier $\bar{\alpha}$. The other, $\gamma $, is determined from the energy
constraint (\ref{Uesc}); using \Ex{rhoT1} in (\ref{Uesc}), we have 
\begin{equation}
0=\frac{\Int \left( 1- \left( 1-q\right)\,{\gamma}\, %
\left( H-U\right)\right) ^{\frac{q}{1-q}}\left(
H-U\right) d\Gamma }{\Int \left( 1- \left( 1-q\right)\,{\gamma}\, %
\left( H-U\right)\right)^{\frac{q}{1-q}}d\Gamma } \,,
\end{equation}%
which by multiplying the numerator by $(1-q)\gamma$ and adding and subtracting one gives
\begin{eqnarray}
{\int \left( 1-\left( 1-q\right) \gamma \,\left( H-U\right) \right) ^{\frac{q%
}{1-q}}d\Gamma } =
{\int \left( 1-\left( 1-q\right) \gamma \,\left( H-U\right)
\right) ^{\frac{1}{1-q}}d\Gamma }\,.  \label{UT}
\end{eqnarray}%
Note that the sign of the factor $f(\Gamma )\equiv 1-\left( 1-q\right)
\gamma \left( H-U\right) $ is, so far, arbitrary so that we allow for the
cancellation of a (possible imaginary) factor throughout these equations.
However, this is only possible in the expression for the distribution
\Ex{rhoT1} if the sign of $f(\Gamma )$ is independent of $\Gamma $. Given
this fact, the sign can be fixed by making the substitution $f(\Gamma
)=s\left|f(\Gamma )\right|$ in the energy constraint which requires that $s^{\frac{q}{%
1-q}}=s^{\frac{1}{1-q}}$ so that $s=1$. Since the kinetic energy is in
principle unbounded, the constant sign of the argument means that the
distribution may have to be restricted to some subset of phase space so that
it should be written as 
%\begin{widetext}
\begin{equation}
\rho \left( \Gamma \right) =\frac{\left( 1- \left( 1-q\right) \gamma %
\left( H-U\right) \right)^{\frac{1}{1-q}} \Theta\left( 1- \left( 1-q\right) \gamma %
\left( H-U\right) \right)}%
{\Int \left( 1-\left( 1-q\right) \gamma%
\left( H-U\right)\right)^{\frac{1}{1-q}} \Theta\left( 1- \left( 1-q\right) \gamma %
\left( H-U\right) \right)d\Gamma}\,,
\label{rhoT11}
\end{equation}%
%\end{widetext}
where $\Theta (x)$ is the step function which is one for $x>0$
and zero otherwise. We note that the introduction of the step function is in fact a redefinition of the variational problem in the sense that we have replaced $\rho(\Gamma)$ with $\bar{\rho}(\Gamma)\Theta(f(\Gamma))$ in Eq.(\ref{STq}-\ref{Uesc}) and then maximized with respect to  $\bar{\rho}(\Gamma)$. If we do not do this, then there is simply no solution to the variational problem which is real and non-negative for all $\Gamma$. 
To see this, we note from the variational equation (\ref{vareq}) 
that if $\rho(\Gamma)$ vanishes for some value of $\Gamma$ then it necessarily follows that $\bar{\alpha} = 0$, which is untenable since $\bar{\alpha}$ will generally assume a non-zero value due to the normalization condition. Therefore, we can only restrict the support of the distribution by redefining the variation problem.
We now turn to the determination of the possible values for~$q$. 

{\it{Case:} $q<1$}. The exponent occuring in the distribution is 
$\frac{1}{1-q}>0$ so that if $\gamma>0$, then the distribution must have finite
support since for some sufficiently large value of $T$, the argument, %
$f(\Gamma)$, becomes zero, and from (\ref{rhoT1}) we have 
\begin{equation}
\rho\left( \Gamma\right) =\frac{\exp_{q}\left( -\gamma\left( H-U\right)
\right) }{\int\exp_{q}\left( -\gamma\left( H-U\right) \right) d\Gamma }\,,
\label{rhoTq}
\end{equation}
where {{$\exp_q(x) \equiv (1+(1-q)x)^{\frac{1}{1-q}}\Theta(1+(1-q)x)$ }}is the 
$q$-exponential function.

If $\gamma<0$, then $f(\Gamma)$ is always positive, and the distribution
function never goes to zero so that there can be no finite support. This
however leads to another problem since $f(\Gamma)$ is unbounded as the
kinetic energy increases which, in turn, means that the integral of $%
f(\Gamma)$ over momenta will diverge so that the distribution cannot be
normalized. We conclude that $\gamma<0$ is not allowed. We remark that it
might be thought that one could introduce a limit on the kinetic energy, but
this is not in keeping with the proposal that the nonextensive distribution
be a generalization of the canonical distribution which describes {\it{open
systems}} in contact with a reservoir.

{\it{Case:} $q>1$}. The exponent being then $\frac{1}{1-q}<0$,
we write the distribution function as%
%\begin{widetext}
\begin{equation}
\rho \left( \Gamma \right) =\frac{\left( 1+ \left\vert 1-q\right\vert \gamma%
 \left( H-U\right) \right) ^{-\left\vert \frac{1}{1-q}\right\vert } %
\Theta \left(1+ \left\vert 1-q\right\vert \gamma%
\left( H-U\right) \right)}%
{\displaystyle \int \left( 1+ \left\vert 1-q\right\vert \gamma \left( H-U\right) \right) ^{-\left\vert \frac{1}{1-q}\right\vert } %
\Theta \left(1+ \left\vert 1-q\right\vert \gamma%
\left( H-U\right) \right) d\Gamma } \,.
\end{equation}%
%\end{widetext}
If $\gamma >0$, the distribution will be normalizable provided
that $\left( \sum p_{i}^{2}\right) ^{-\left\vert \frac{1}{1-q}\right\vert }$
is integrable over $d^{ND}p$ (for large $p$) which is to say that $\left(
P^{2}\right) ^{-\left\vert \frac{1}{1-q}\right\vert }P^{ND-1}dP$ be
integrable for $P\rightarrow \infty $; this means we must have $%
-1>ND-1-2\left\vert \frac{1}{1-q}\right\vert $, 
%\begin{equation}
or $1<q<1+\frac{2}{ND}$ 
%\label{q>1}
%\end{equation}%
which condition reduces to the classical Boltzmann result ($q=1$) 
in the thermodynamic limit~\footnote{Note that $\rho^{q}\left( \Gamma\right) $ must also be
integrable which imposes that $-1 > ND-1-\frac{2q}{q-1}$ or $q < 1+\frac {2}{%
ND-2}$, but this condition is weaker than the constraint 
$1<q<1+\frac{2}{ND}$.}.
 
%\begin{equation}
%1+{\gamma \left\vert 1-q\right\vert }(V-U)>0\,.
%\end{equation}%
The other possibility is $\gamma <0$. We then write the distribution as%
%\begin{widetext}
\begin{equation}
\rho \left( \Gamma \right) =\frac{\left( 1-\left\vert \gamma \right\vert
\left\vert 1-q\right\vert \left( H-U\right) \right)%
^{-\left\vert \frac{1}{1-q}\right\vert} %
\Theta \left( 1-\left\vert \gamma \right\vert \left\vert 1-q\right\vert
\left( H-U\right) \right) }%
{\displaystyle \int \left( 1-\left\vert \gamma \right\vert \left\vert
1-q\right\vert \left( H-U\right) %
\right) ^{-\left\vert \frac{1}{1-q}\right\vert }\Theta
\left( 1-\left\vert \gamma \right\vert \left\vert 1-q\right\vert %
\left( H-U\right) \right)
d\Gamma }
\end{equation}%
%\end{widetext}
When integrated over momenta, this expression would show a
singularity at %\begin{equation}
$T=\frac{1}{\left\vert \gamma \right\vert \left\vert 1-q\right\vert }+U-V$, 
%\end{equation}%
unless \\ %\begin{eqnarray}
$\int {\left( X-\sum_{i}p_{i}^{2}\right) ^{-\left\vert \frac{1}{1-q}\right\vert }}dp^{3N} %\\
 \sim \int_{0}^{\sqrt{X}}{\left( X-P^{2}\right)^{\left\vert \frac{-1}{1-q}\right\vert }}P^{3ND-1}dP %\\
\sim \int_{0}^{X}{\left(X-Y\right) ^{\left\vert \frac{-1}{1-q}\right\vert }}Y^{\frac{3ND-2}{2}}dY$
%\end{eqnarray}%
is finite; this requires that $1-\left\vert \frac{1}{1-q}\right\vert =1-%
\frac{1}{q-1}>0$, i.e. $q>2$ . But $\rho ^{q}\left( \Gamma \right) $ must
also be integrable and in order that the singularity be integrable imposes $%
1-\frac{q}{q-1}>0$ which is incompatible  with the  condition $q>2$.

Furthermore the maximization condition demands that the second derivative of the 
$q$-entropy \Ex{STq} be $- k_B\,K\, q \,\rho ^{q - 2} < 0$, which is satisfied if $q > 0$.
So in summary, the distribution function exists  in the thermodynamic limit for $\gamma> 0$ 
and  $0\leq q \leq 1$, and  $\rho\left( \Gamma\right) $ has the form of a $q$-exponential 
with finite support.
%while in the other possible case ($1 < q < 1+\frac{2}{ND}$) it
%reduces to the classical Boltzmann result ($q=1$) in the thermodynamic limit.

\section{Homogeneous entropy}

The homogeneous entropy was proposed as an alternative to the Tsallis entropy because it has various desirable properties that the Tsallis entropy does not share such as being Lesche stable and giving positive-definite heat capacities \cite{lutsko_boon_grosfils}. It is therefore interesting to ask whether it is subject to the same limitations as found for the Tsallis entropy. The analysis follows essentially the same lines as above except that in this case $\rho\left(\Gamma\right) $ is the physical probability~\cite{lutsko_boon_grosfils} and
so the energy is computed with the normal average. The homogeneous entropy
reads~\cite{lutsko_boon_grosfils} 
\begin{equation}
S_{q}^{H}=k_{B}\,\frac{1-\left( K\int\rho^{\frac{1}{q}}\left( \Gamma\right)
d\Gamma\right) ^{q}}{1-q}\,,  
\label{SHq}
\end{equation}
where $K$ is a quantity with the dimensions $\left[ \Gamma\right] ^{\frac{1-q%
}{q}}$, i.e. $K=\hbar^{DN\,\frac{1-q}{q}}$, and the normalization and energy
constraints are 
\begin{equation*}
1=\int\rho\left( \Gamma\right) d\Gamma\;\;\;\;;\;\;\;\;U={\int\rho\left(
\Gamma\right) H\,d\Gamma}\,.
\end{equation*}
Following a similar analysis as given in the previous section, %\cite {epaps},
the conclusions are that the condition for normalizability is $0\leq q\leq1$ and 
$\gamma>0$ in which case the distribution function reads 

\begin{equation}
\rho\left( \Gamma\right) \,=\,\frac{\left( \exp_{q}\left( -\,\gamma \,\left(
H-U\right) \right) \right) ^{q}}{\int\,d\Gamma\,\left( \exp _{q}\left(
\,-\gamma\,\left( H-U\right) \right) \right) ^{q}}\;. 
 \label{rhoHq} 
\end{equation}

Here
$\gamma = \frac{\bar{\gamma}}{\left(K \int \rho ^{\frac{1}{q}} \left( \Gamma \right) d\Gamma \right)^q}$
where $\bar{\gamma}$ is the Lagrange multiplier used to fix the average energy
and $\rho\left( \Gamma\right)$ has finite support. 

\section{Thermodynamic properties}

Given that the formalism is constructed 
solely on the basis of the three axioms \Ex {STq}, \Ex{Norm} and \Ex{Uesc},  
the consistent way to define the thermodynamic temperature is through the 
thermodynamic definition ${\partial S}/{\partial U}= 1/T$. 
Considering $q<1$ and  $\gamma >0$, 
we obtain %(see \cite{epaps} for details)
from the Tsallis entropy~\Ex{STq} 
\begin{eqnarray}
{T^T_q}=\frac{1}{ k_B\,\gamma}%
\left( K^{\frac{1}{1-q}} \int   \exp_q\left(-\gamma%
\left(H-U\right)\right) d\Gamma \right)^{q-1} \,,
\label{TqT}
\end{eqnarray}%
and from the homogeneous entropy~\Ex{SHq}
\begin{eqnarray}
T^H_q \,=\,\frac{q}{k_B\,\gamma}
\left( K^{\frac{q}{q-1}} \int \exp_q \left( - \frac{\gamma}{q}\left(H-U\,\right)\right)%
d\Gamma \right)^{1-q}  \label{THq}
\end{eqnarray}
With the explicit expressions of ${\gamma}$ (see \Ex{rhoTq} and \Ex{rhoHq}), \Ex{THq}
gives ${\bar{\gamma}}\,=\,{1}/{(k_B\,T^H_q)}$, the analog of the classical expression, whereas
the equivalent relation for the Tsallis temperature is only obtained in the limit $q \rightarrow 1$.
%${\bar{\gamma}}\,=\,{1}/{(k_B\,T^T_{q=1})}\,=\,{1}/{(k_B\,T)}$. 
 
Correspondingly, the expressions for the specific heat  $C_{V}=\left( \frac{\partial U}{\partial T_q}\right)$ 
are given by %\cite{epaps}
\begin{eqnarray}
C^T_{V}=  
\frac{{\beta^T}/{\gamma}}{\left[ \frac{1}{q}\left(\frac{\beta^T}{\gamma}\right)^{4}\frac{1}{K^{2}%
\int\rho^{2q-1}\left(  \Gamma\right)  \left(  \beta^T\left(  H-U\right)
\right)^{2}d\Gamma}-2\left(  1-q\right)  \right]}  %^{-1}
\end{eqnarray}
with the classical notation $\beta^T = 1/(k_BT^T_q)$, and by
\begin{equation}
C^H_{V} =  q^{\frac{1}{q}} \left( {\frac{\beta^H}{\gamma}} \right)^{\frac{1-q}{q}} \, {\frac{k_{B}}{K}}%
\int\rho^{\frac{2q-1}{q}}\left(  \Gamma\right)  \left(  \beta^H\left(H-U\right)  \right)  ^{2}d\Gamma
\label{CvH}
\end{equation}
where $\beta^H = 1/(k_BT^H_q)$, or 
%Interestingly \Ex{CvH} can also be rewritten as
\begin{equation}
C^H_{V} =  {\frac{1}{k_{B}\,(T_q^H)^2}}%
\int d\Gamma \rho \left(  \Gamma\right) \left(H-U\right)^{2}\,\mathsf{C}_q(\Gamma)
\label{CvH1}
\end{equation}
with $\mathsf{C}_q(\Gamma) = q^{\frac{1}{q}}\left(\frac{\beta^H}{\gamma}\right)^{\frac{1-q}{q}}%
\frac{\rho^{\frac{q-1}{q}}\left(  \Gamma\right)}{K}$. \Ex{CvH1} is the generalization of the classical 
expression of the specific heat given in terms of the energy fluctuations:
$C_{V} =  {\langle (\Delta E)^2\rangle}/({k_{B}\,T^2})$.
The thermodynamic temperatures are both positive while the generalized specific heat is only 
positive-definite in the case of the homogeneous entropy.\footnote{{It was indeed shown that in the 
Tsallis formulation the specific heat can be negative~\cite{abe_99}.}}

\section{Concluding comments}

Non-exponential distributions are widely observed in nature. Non-extensive
thermodynamics was motivated, in part, as a means of explaining the origin
of such distributions which arise naturally as a result of maximizing the
generalized entropy with the usual constraints of the normalization of the
distribution and of fixed average energy. We have shown that when this
procedure is applied to Hamiltonian systems, the resulting distributions
only exist for the restricted range of $0<q<1+{\mathcal O}\left( \frac{1}{N}\right) $.
Since the so-called "fat-tailed" distributions correspond to $q>1$, this
means that generalized thermodynamics cannot be seen as an explanation of
their occurrence for these systems. The problem with larger values of $q$ has
to do with the existence of the integrals over momenta due to the
unboundedness of the kinetic energy. One way around this would be to
redefine the formalism so as to restrict the momenta a priori by making the
ansatz ${\rho\left( \Gamma\right) =\Theta}\left( T_{\ast}-T\right) {\rho}%
_{\ast}{\left( \Gamma\right) }$, for some fixed positive number $T_{\ast}$,
throughout the variational problem and maximizing to determine ${\rho}_{\ast}%
{\left( \Gamma\right) }$. However, this is obviously quite artificial and
{\it ad hoc} since, for example, one could replace the step function by any
function of $T$ that goes to zero sufficiently quickly as $T$ grows. 
%\remove{Even for the case $0<q<1$ a similar problem exists since we find that the range in phase space must also be restricted and again there is no reason that the restriction must be maximal, as is usually assumed. In other words, instead of $\rho(\Gamma)\Theta(a(\Gamma))$ we could have used $\rho(\Gamma)\Theta(a(\Gamma)-(1-q)\epsilon)$ for any $\epsilon>0$ since the only necessary property is that this go to the Boltzmann distribution in the $q \rightarrow 1$ limit. However, the choice $\epsilon=0$ \it{is} necessary in order for the distribution to be differentiable at all points as is required in order to derive the thermodynamic temperature.}  
This suggests the more straightforward conclusion that in the case of classical
Hamiltonian systems, nonextensive thermodynamics does not provide a simple,
natural explanation of distributions with fat tails.

\acknowledgments
This work was {partly} supported by the European Space Agency under contract
number ESA AO-2004-070.


\bigskip


\begin{thebibliography}{99}

\bibitem{davis} A. Scott, ed.,  {\textit {Encyclopedia of Nonlinear Science}}
(Taylor and Francis, New York, 2005).

\bibitem{tsallis88} C. Tsallis, \textit{J. Stat. Phys.}, \textbf{52}, 479
(1988).

%\bibitem{footnote0} We note that other approaches exist such as super-statistics
%\cite{beck-cohen} and the fractional and nonlinear Fokker-Planck equations 
%(see e.g. \cite{lutsko-boon} and references therein).

\bibitem{beck-cohen} C. Beck and E.G.D. Cohen, \textit{Physica A}, \textbf{321}, 267 (2003).
  
\bibitem{lutsko-boon} J. F. Lutsko and J. P. Boon, \textit{Phys. Rev. E}, \textbf{77}, 051103 (2008).

%\bibitem{gellmann-tsallis} M. Gell-Mann and C. Tsallis, \it{Nonextensive
%Entropy} (Oxford University Press, New York, 2004).

\bibitem{swinney-tsallis} H.L. Swinney and C. Tsallis, eds., \textit {Anomalous
Distributions, Nonlinear Dynamics, and Nonextensivity}, \textit{Physica D}, 
\textbf{193} (2004); J.P. Boon and C. Tsallis, eds.,  \textit {Nonextensive
statistical mechanics: new trends, new perspectives}, \textit{Europhys. News}%
, \textbf{36/6}, {183-231} {(2005)}; Chap.7 in~\cite{tsallis09}.

\bibitem{tsallis09} {C. Tsallis},  \textit {Introduction to Nonextensive
Statistical Mechanics} {(Springer, New York, 2009)}, see Bibliography.

\bibitem{opponents} R. Luzzi, A.R. Vasconcelos and J. Galvao Ramos, \textit{%
Science}, \textbf{298}, 1171 (2002); M. Nauenberg, \textit{Phys. Rev. E}, 
\textbf{67}, 036114 (2003); D.H. Zanette and M.A. Montemurro, \textit{%
Physics Lett. A}, \textbf{316}, 184 (2003); P. Grasberger, \textit{Phys.
Rev. Lett.}, \textbf{95}, 140601 (2005); D.H. Zanette and M.A. Montemurro, 
\textit{Physics Lett. A}, \textbf{324}, 48 (2007).

\bibitem{conflict} A. Cho, \textit{Science}, \textbf{297}, 1269 (2002); A.
Plastino, \textit{Science}, \textbf{300}, 250 (2003); V. Latora, A.
Rapisarda and A. Robledo, \textit{Science}, \textbf{300}, 250 (2003); C.
Tsallis, \textit{Phys. Rev. E}, \textbf{69}, 038101 (2004); M. Nauenberg, 
\textit{Phys. Rev. E}, \textbf{69}, 038102 (2004); R. Balian and M.
Nauenberg, \textit{Europhys. News}, \textbf{37}, {9} {(2006)}; F. Bouchet,
T. Dauxois and S. Ruffo, \textit{Europhys. News}, \textbf{37}, {9} {(2006)};
A. Rapisarda and A. Pluchino, \textit{Europhys. News}, \textbf{37}, {10} {%
(2006)}; R. Luzzi, A.R. Vasconcelos and J. Galvao Ramos, \textit{Europhys.
News}, \textbf{37}, {11} {(2006)}; see also discussion in Chap.8 of
reference~\cite{tsallis09}.

\bibitem{abe_10} {S. Abe}, \textit{Europhys. Lett.}, \textbf{90}, {50004} {(2010)}.

\bibitem{abe_99} 
{{{S. Abe}, \textit{Phys. Lett. A}  \textbf{263} {424}  {(1999)}; \textbf{267} {456} {(2000)}.}}

\bibitem{boon_lutsko_11}
{J.P. Boon and J.F. Lutsko}, \textit{Physics Lett. A}, \textbf{375}, {329}  {(2011)}.

\bibitem{lesche_abe} 
{{{B. Lesche}, \textit{J. Stat. Phys.} \textbf{27} {419} {(1982)};
{S. Abe}, \textit{Phys. Rev. E} \textbf{66} {046134} {(2002)}.}}

\bibitem{lutsko_boon_grosfils} {J.F. Lutsko, J.P. Boon and P. Grosfils}, 
\textit{Europhys. Lett.}, \textbf{86}, {40005} {(2009)}.

\bibitem{beck_schlogl} C. Beck and F. Schl\"ogl, {\textit {Thermodynamics of
Chaotic Systems}}, (Cambridge University Press, London, 1993).

%\bibitem{footnote1} {For simplicity we omit the Boltzmann constant $k_{B}$
%which will be reintroduced when necessary.}

%\bibitem{footnote2} {Note that $\rho^{q}\left( \Gamma\right) $ must also be
%integrable which imposes that $-1 > ND-1-\frac{2q}{q-1}$ or $q < 1+\frac {2}{%
%ND-2}$, but this condition is weaker than the constraint in the text.} %(\ref{q>1}).}




\end{thebibliography}
\end{document}